\documentclass[twoside,twocolumn,9pt]{article}
\usepackage{extsizes}
\usepackage[super,sort&compress,comma]{natbib}
\usepackage[version=3]{mhchem}
\usepackage[left=1.5cm, right=1.5cm, top=1.785cm, bottom=2.0cm]{geometry}
\usepackage{balance}
\usepackage{mathptmx}
\usepackage{sectsty}
\usepackage{graphicx}
\usepackage{lastpage}
\usepackage[format=plain,justification=justified,singlelinecheck=false,font={stretch=1.125,small,sf},labelfont=bf,labelsep=space]{caption}
\usepackage{float}
\usepackage{fancyhdr}
\usepackage{fnpos}
\usepackage[english]{babel}
\addto{\captionsenglish}{%
  
}
\usepackage{array}
\usepackage{droidsans}
\usepackage{charter}
\usepackage[T1]{fontenc}
\usepackage[usenames,dvipsnames]{xcolor}
\usepackage{setspace}
\usepackage[compact]{titlesec}
\usepackage{hyperref}

\usepackage{epstopdf}

\usepackage{soul} 
\usepackage{units} 
\definecolor{cream}{RGB}{222,217,201}

\begin{document}

\pagestyle{fancy}
\thispagestyle{plain}
\fancypagestyle{plain}{
\renewcommand{\headrulewidth}{0pt}
}

\makeFNbottom
\makeatletter
\renewcommand\LARGE{\@setfontsize\LARGE{15pt}{17}}
\renewcommand\Large{\@setfontsize\Large{12pt}{14}}
\renewcommand\large{\@setfontsize\large{10pt}{12}}
\renewcommand\footnotesize{\@setfontsize\footnotesize{7pt}{10}}
\makeatother

\renewcommand{\thefootnote}{\fnsymbol{footnote}}
\renewcommand\footnoterule{\vspace*{1pt}%
\color{cream}\hrule width 3.5in height 0.4pt \color{black}\vspace*{5pt}}
\setcounter{secnumdepth}{5}

\makeatletter
\renewcommand\@biblabel[1]{#1}
\renewcommand\@makefntext[1]%
{\noindent\makebox[0pt][r]{\@thefnmark\,}#1}
\makeatother
\renewcommand{\figurename}{\small{Fig.}~}
\sectionfont{\sffamily\Large}
\subsectionfont{\normalsize}
\subsubsectionfont{\bf}
\setstretch{1.125} 
\setlength{\skip\footins}{0.8cm}
\setlength{\footnotesep}{0.25cm}
\setlength{\jot}{10pt}
\titlespacing*{\section}{0pt}{4pt}{4pt}
\titlespacing*{\subsection}{0pt}{15pt}{1pt}

\fancyfoot{}
\fancyfoot[LO,RE]{\vspace{-7.1pt}\includegraphics[height=9pt]{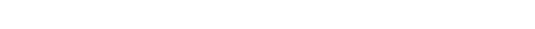}}
\fancyfoot[CO]{\vspace{-7.1pt}\hspace{11.9cm}\includegraphics{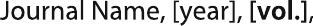}}
\fancyfoot[CE]{\vspace{-7.2pt}\hspace{-13.2cm}\includegraphics{head_foot/RF}}
\fancyfoot[RO]{\footnotesize{\sffamily{1--\pageref{LastPage} ~\textbar  \hspace{2pt}\thepage}}}
\fancyfoot[LE]{\footnotesize{\sffamily{\thepage~\textbar\hspace{4.65cm} 1--\pageref{LastPage}}}}
\fancyhead{}
\renewcommand{\headrulewidth}{0pt}
\renewcommand{\footrulewidth}{0pt}
\setlength{\arrayrulewidth}{1pt}
\setlength{\columnsep}{6.5mm}
\setlength\bibsep{1pt}

\makeatletter
\newlength{\figrulesep}
\setlength{\figrulesep}{0.5\textfloatsep}

\newcommand{\topfigrule}{\vspace*{-1pt}%
\noindent{\color{cream}\rule[-\figrulesep]{\columnwidth}{1.5pt}} }

\newcommand{\botfigrule}{\vspace*{-2pt}%
\noindent{\color{cream}\rule[\figrulesep]{\columnwidth}{1.5pt}} }

\newcommand{\dblfigrule}{\vspace*{-1pt}%
\noindent{\color{cream}\rule[-\figrulesep]{\textwidth}{1.5pt}} }

\makeatother

\twocolumn[
  \begin{@twocolumnfalse}
{\includegraphics[height=30pt]{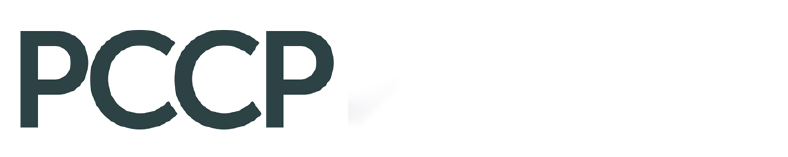}\hfill\raisebox{0pt}[0pt][0pt]{\includegraphics[height=55pt]{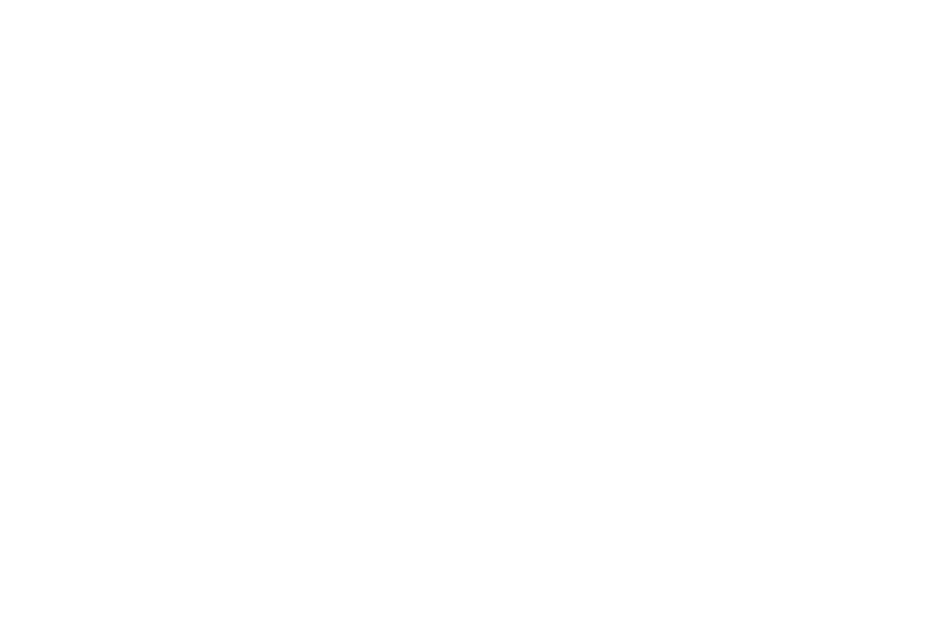}}\\[1ex]
\includegraphics[width=18.5cm]{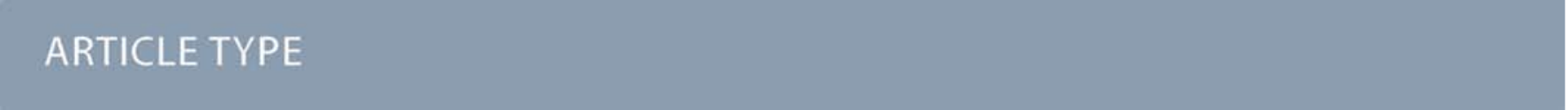}}\par
\vspace{1em}
\sffamily
\begin{tabular}{m{4.5cm} p{13.5cm} }

\includegraphics{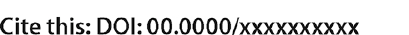} & \noindent\LARGE{\textbf{The Li + CaF $\longrightarrow$ Ca + LiF chemical reaction under cold conditions}} \\
\vspace{0.3cm} & \vspace{0.3cm} \\

& \noindent\large{Humberto da Silva Jr.\textit{$^{a}$}, Qian Yao\textit{$^{b}$}, Masato Morita\textit{$^{c}$}, Brian K. Kendrick\textit{$^{d}$}, Hua Guo\textit{$^{b}\dag$} and Naduvalath Balakrishnan\textit{$^{a}\ddag$}} \\ 

\includegraphics{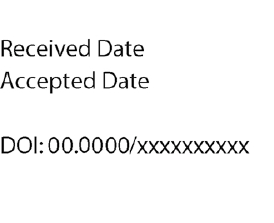} & \noindent\normalsize{The calcium monofluoride (CaF) molecule has emerged as a promising candidate for precision measurements, quantum simulation, and ultracold chemistry experiments. Inelastic and reactive collisions of laser cooled CaF molecules in  optical tweezers have recently been reported and collisions of cold Li atoms with CaF are of current experimental interest. In this paper, we report ab initio electronic structure and full-dimensional quantum dynamical calculations of the Li + CaF $\to$ LiF + Ca chemical reaction. The electronic structure calculations are performed using the internally contracted multi-reference configuration-interaction method with Davidson correction (MRCI+Q). An analytic fit of the interaction energies is obtained using a many-body expansion method. A coupled-channel quantum reactive scattering approach implemented in hyperspherical coordinates is adopted for the scattering calculations under cold conditions. Results show that the Li + CaF reaction populates several low-lying vibrational levels and many rotational levels of the product LiF molecule and that the reaction is inefficient in the 1-100 mK regime allowing sympathetic cooling of CaF by collisions with cold Li atoms.} \\

\end{tabular}

 \end{@twocolumnfalse} \vspace{0.6cm}

  ]

\renewcommand*\rmdefault{bch}\normalfont\upshape
\rmfamily
\section*{}
\vspace{-1cm}


\footnotetext{\textit{$^{a}$~Department of Chemistry and Biochemistry, University of Nevada,
Las Vegas, Nevada 89154, USA.}}
\footnotetext{\textit{$^{b}$~Department of Chemistry and Chemical Biology, University of
New Mexico, Albuquerque, New Mexico 87131, USA.}}
\footnotetext{\textit{$^{c}$~Chemical Physics Theory Group, Department of Chemistry and
Center for Quantum Information and Quantum Control, University of
Toronto, Toronto, Ontario M5S 3H6, Canada.}}
\footnotetext{\textit{$^{d}$~Theoretical Division (T-1, MS B221), Los Alamos National Laboratory,
Los Alamos, NM 87545, USA.}}


\footnotetext{\dag~E-mail: hguo@unm.edu}
\footnotetext{\ddag~E-mail: naduvala@unlv.nevada.edu}



\section{Introduction}
The rich internal structure of ultracold molecules
compared to ultracold atoms lend themselves to many applications in emerging areas of quantum science.
Ultracold paramagnetic molecules such as Calcium monofluoride,
CaF, whose electronic ground state is characterized by a $^{2}\Sigma^{+}$
term, have long been considered as a promising candidate for a number
of applications, in particular, quantum simulation \cite{micheli2006,micheli2007,Pupillo_chapter,blackmore2018}, quantum information \cite{demille2002,yelin2006,yelin2009,sawant2020,burchesky2021},
and precision spectroscopy \cite{carr2009}.
This is mostly due to the presence of an unpaired electron
as its resultant non-zero electric and magnetic moment serves as a
convenient experimental handle for extra control \cite{dirosa2004,krems2004b,maussang2005},
by means of external fields (\textit{e.g.} Stark and Zeeman effects).
Additionally, these systems also
provide a unique opportunity to improve upon the fundamental understanding
of atom-molecule \cite{jurgilas2021} and molecule-molecule interactions \cite{miranda2011}, dipolar
interactions \cite{ni2010,ospelkaus2010a} and collision-induced chemistry at the ultra-low range of kinetic
energies \cite{ospelkaus2010b,ospelkaus2010c,singh2012,hu2019,liu2021}.
In particular, experimental explorations of collision-induced
trap-loss rate of molecules with singlet and triplet spin multiplicities,
in  ultracold conditions, have been available for a while (for
systems such as Rb$_{2}$, NaRb, KRb, CsRb, NaK, LiNa, NaRb). However, such
studies are less prevalent for doublet molecules \cite{cheuk2020}.

Slowing the translational motion of CaF molecules down to the capture
velocity of a 800 mK deep magneto-optical trap (MOT) has been recently
achieved by Doyle and co-workers \cite{lu2014}. This follows similar success with SrF, to our knowledge, the first such molecule to be trapped in a MOT~\cite{barry2014,mccarron2015,norrgard2016,steinecker2016}.
The original work of Lu \textit{et al.}, since then improved to sub-Doppler temperatures ~\cite{yeo2015,hemmerling2016,truppe2017,chae2017,anderegg2017},
represents an important milestone after the seminal work of Di Rosa
\cite{dirosa2004}, the first to observe that molecules such as CaF,
CaH, CaOH, SrF, SrOH, YbF, may possess a rovibrational internal structure
with a large one-photon oscillator strength and highly diagonal Franck-Condon
factors. This, in turn, unlocks the possibility of light-assisted
closed cycling transitions, similar to the laser cooling techniques
applied to atoms and atomic ions \cite{anderegg2018}.

Once CaF molecules in the electronic ground state are properly
trapped, as demonstrated by Lu \textit{et al.}, a natural next step
is the design and implementation of cold collisions between
CaF molecules and, say, co-trapped laser cooled atoms or another CaF molecule.
The latter case has been recently realized in a pioneering experiment,
in which  CaF molecules are loaded from a MOT into optical
tweezers and, by varying the relative position of  two tweezers,
CaF + CaF ultracold collisions have been observed to produce two-body
loss, most likely due to yet undetermined chemical reactions, with
magnitude comparable to a theoretical universal loss rate \cite{anderegg2019,cheuk2020}.
The former case of CaF collisions with laser cooled atoms is yet an open prospect and, as we shall see below,
one of the underlying motivation of the present work. Among the Alkali
metal candidates, whose laser cooling and trapping techniques are
nowadays routine procedures, only Li$\left(^{2}S\right)$ combined
with the electronic ground state of CaF provides an exothermicity
of about -4440 cm$^{-1}$ \cite{kosicki2017}. Other atomic species
such as Na, K, Rb and Cs would require a few thousand wavenumber of
collision-induced excitation in order to trigger chemical events \cite{kosicki2017}. However, due
to the low ($<$ 1 K) kinetic energies involved in such experiments,
in general, these collision-induced excitation are all but forbidden
energetically.

Thus, the prospects of Li$\left(^{2}S\right)$ + CaF$\left(X^{2}\Sigma^{+}\right)$
ultracold reactive collisions are highly regarded as an opportunity
to study cold chemistry as well as collision-induced trap losses due to
chemical events. To understand and to establish the limits for sympathetic
cooling of CaF$\left(X^{2}\Sigma^{+}\right)$ toward even lower temperatures
by means of soft collisions with a Li$\left(^{2}S\right)$ coolant
buffer \cite{lim2015} as well as collisional shielding \cite{quemener2016,karman2019} a detailed investigation of Li + CaF collisions is needed. Prior studies of Li$\left(^{2}S\right)$ + CaF$\left(X^{2}\Sigma^{+}\right)$ collisions explored only elastic and (non-reactive) inelastic collisions using model potentials or interaction potentials that do not describe the reactive regions. Foreseeing an upcoming demand for more theoretical support
in regard to this matter, in this work, we tackle the challenging
task of describing a new LiCaF global potential energy surface (PES)
and to perform the first description of the Li$\left(^{2}S\right)$ + CaF$\left(X^{2}\Sigma^{+}\right)$ $\longrightarrow$
Ca$\left(^{1}S\right)$ + LiF$\left(X^{1}\Sigma^{+}\right)$ collisions resorting to state-of-the-art quantum reactive scattering, \textit{\textcolor{black}{i.e.}} a coupled-channel (CC) method. It is worthwhile to note that a novel full six-dimensional PES intended to describe the even more complicated CaF + CaF $\longrightarrow$ CaF$_{2}$ + Ca chemical reaction has been constructed by Sardar and co-workers \cite{sardar2022}.

Until very recently a proper quantum description of the title reaction was not feasible. Today,
by employing unprecedented computational resources, it remains
a very hard numerical task due to several reasons, namely: (i) the system
lacks symmetries that could otherwise be used to ease parts of the
computational overload; (ii) it is a somewhat heavy system with small
diatomic rotation constants (\textit{e.g.} the CaF constant is about
177 times smaller than that of H$_{2}$) and, as we shall see below,
possesses a relatively deep potential well at short range, all of
which translates into a large amount of spatially delocalized internal states required to properly
describe the collision; (iii) it is known to be a very anisotropic
system characterized by strong couplings between collisional channels
that would be negligible otherwise; and (iv),
typical of atom-molecule collisions within the cold domain of
kinetic energies, the radial solution of the Schr$\ddot{\mathrm{o}}$dinger equation
is required to be propagated to unusually large atom-molecule separations, due in part
to the extremely long de Broglie wavelength of the colliding partners. Therefore,
within the limitations imposed by such aspects, we provide below a
first investigation on the optimal parameters required to extract accurate
scattering characteristics for these collisions, in a time-independent
quantum reactive scattering formalism, and discuss the predicted features
of the collisional cross sections as  functions of the incident
energy. To this end, the adiabatically adjusting
principal axis hyperspherical (APH) quantum reactive scattering suite of
programs (hereafter referred to as APH3D), that has been used to describe a diverse array of reactive
collisional problems in our group \cite{pack1987,kendrick1999,kendrick2000,kendrick2001,kendrick2016a,kendrick2003a,kendrick2003b,kendrick2016b,kendrick2018a,kendrick2018b,kendrick2019}, is also utilized below. While formalisms based on the solution of the time-dependent Schr{\"o}dinger equation are computationally more efficient they are slow to converge at low collision energies~\cite{Huang2018,Huang2021}. Methods based upon statistical quantum approaches~\cite{Rakham2003,Makrides2015,Yang2020_SQM} have also been applied to complex-forming ultracold chemical reactions but their accuracy for state-to-state transitions is not fully established.

The paper is organized as follows: Section \ref{sec:theory} provides a brief description of the theoretical approach with details of the electronic structure calculations presented in subsection \ref{subsec:pes}. A brief outline of the quantum scattering formalism using the APH3D code is presented in subsection \ref{subsec:aph3d}. Section \ref{sec:results} presents the results and section \ref{sec:conclusions} provides a summary of our findings.

\section{Theoretical Approach}
\label{sec:theory}

\subsection{Potential energy surface}
\label{subsec:pes}

The Li$\left(^{2}S\right)$ + CaF$\left(X^{2}\Sigma^{+}\right)$
reactants asymptotically correlate with the triatomic electronic
states $^{1}A^{\prime}$ and $^{3}A^{\prime}$. In what follows we describe the computation of the ground electronic state, $X^{1}$A$^{\prime}$, of the LiCaF complex
using the internally contracted multi-reference
configuration-interaction method with the Davidson correction (MRCI+Q)
\cite{shiozaki2011a,shiozaki2011b,shiozaki2013}, as implemented in the MOLPRO package \cite{werner2008}.
The augmented
correlation-consistent polarized valence quadruple-zeta basis set (aug-cc-pVQZ)
of Dunning \cite{prascher2011,dunning1989,kendall1992} was used for the Li and F atoms, whereas the cc-pwCVQZ-PP basis, in which the core electrons are described
with a pseudopotential, was used for the Ca atom \cite{hill2017}. Calculations
with full valence active space utilizing a state-averaged ($1^{1}$A$^{\prime}$,
$1^{3}$A$^{\prime}$ and $1^{1}$A$^{\prime\prime}$) complete active
space (10 active electrons in 9 active orbitals) self-consistent field
wavefunction (SA-CASSCF) \cite{werner1985,knowles1985}
were performed. The active space included the 2$s$, 2$s$2$p$, 4$s$4$p$ orbitals from Li, F, and
Ca atoms, whereas the 1$s$ orbitals for Li and F, along with the
3$s$3$p$ orbitals of Ca were closed in the CASSCF calculations and
further cored in the MRCI calculations.

A total of about 11000 geometries below 4.5 eV relative to the global
minimum were selected and fitted using a many-body expansion method \cite{murrel1984b}

\begin{equation}
V_{abc}\left(r_{ab}, r_{ac}, r_{bc}\right)=\sum_{a}V_{a}^{(1)} + \sum_{ab}V_{ab}^{(2)}\left(r_{ab}\right) + V_{abc}^{(3)}\left(r_{ab}, r_{ac}, r_{bc}\right),
\end{equation}

\noindent in which $r_{xy}$ is the internuclear distance between $x$ and $y$ (= $a$, $b$, or $c$); $V_{a}^{(1)}$, $V_{ab}^{(2)}$ and $V_{abc}^{(3)}$ are the one-, two-, and three-body terms, respectively. The one-body terms in Eq. (1) are set to zero. The two-body terms correspond to the diatomic potential energy curves (PECs). The three-body energy becomes zero at all the dissociation limits.

The two-body terms, $V_{\mathrm{CaF}}^{(2)}$ and $V_{\mathrm{LiF}}^{(2)}$, are spline-interpolated in the ranges of 3.2 \textit{a.u.} $\le r_{\mathrm{CaF}} \le$ 7 \textit{a.u.} and 2.4 \textit{a.u.} $\le  r_{\mathrm{LiF}} \le$ 5.6 \textit{a.u.}, respectively. Outside the interpolated regions, the PECs are approximated by the Morse form,

\begin{equation}
V_{\mathrm{morse}}^{(2)}\left(r_{xy}\right)=D_{e}\left[\left(1 - e^{-\alpha_{xy}\left(r_{xy}-r_{e}\right)}\right)^{2}-1\right],
\end{equation}

\noindent where $D_{e}$ is the dissociation energy, $r_{e}$ is the equilibrium distance of the diatoms, and $\alpha$ is a parameter. In the CaF case, $\left(D_{e}=5.45\,\mathrm{eV},\alpha=0.51\,a.u.^{-1},r_{e}=3.92\,a.u.\right)$ are used for $r<3.2$ \textit{a.u.}, whereas $\left(D_{e}=5.45\,\mathrm{eV},\alpha=0.44\,a.u.^{-1},r_{e}=3.4\,a.u.\right)$ are for $r>7$ \textit{a.u.} Similarly, in the LiF case, $\left(D_{e}=5.95\,\mathrm{eV},\alpha=0.47\,a.u.^{-1},r_{e}=3.23\,a.u.\right)$ for $r<2.4$ \textit{a.u.} and $\left(D_{e}=5.95\,\mathrm{eV},\alpha=0.38\,a.u.^{-1},r_{e}=2.62\,a.u.\right)$ for $r>5.6$ \textit{a.u.}

The three-body term is expressed as a polynomial of order M,

\begin{equation}
V_{abc}^{(3)}\left(r_{ab}, r_{ac}, r_{bc}\right)=\sum_{jkl}^{M}d_{jkl}\rho_{ab}^{j}\rho_{ac}^{k}\rho_{bc}^{l},
\end{equation}

\noindent where $\rho_{xy}=r_{xy}e^{-\beta_{xy}r_{xy}}$. The linear parameters, $d_{jkl}$, can be obtained by the linear least squared method and the nonlinear parameters, $\beta_{xy}$, are set to 0.5 \textit{a.u.}$^{-1}$. Moreover, the constraints $j+k+l \neq j \neq k \neq l$ and $j+k+l \le M$ are employed to ensure the three-body term $V_{abc}^{(3)}$ is going to zero at all dissociation limits. In this work, the value of $M=8$ is used, which leads to a total of 140 $d_{jkl}$ linear coefficients. The root mean squared deviation (RMSE) of the three-body short-range fit is 22.7 meV.

The \textit{ab initio} calculation yielded an exothermicity of -0.37 eV (-2984.2 cm$^{-1}$) for the  Li$\left(^{2}S\right)$ + CaF$\left(X^{2}\Sigma^{+}\right)$ reaction, which is 0.13 eV (1048.5 cm$^{-1}$) higher than the experimental value of -0.5 eV (-4033 cm$^{-1}$). This error is corrected in the two-body terms which are adjusted to reproduce the experimental exothermicity.

The long-range interaction potential, $V_{\mathrm{LR}}$, in each arrangement is fitted with the following expression:

\begin{equation}
V_{\mathrm{LR}}=\sum_{nml}C_{nml}r^{l}\frac{B_{n}^{m}\left(\theta\right)}{R^{n}},
\end{equation}

\noindent where $V_{LR}=V_{abc}-V^{(1)}_{a}-V^{(2)}_{bc}$ and $R$ is the distance
between the Li (Ca) atom and the center of mass of the CaF (LiF) molecule. The parameters $l$ and $n$ range from -3 to 3 and 4 to 7, respectively.
For $n=4$ and $m=1$, $B_{4}^{1}\left(\theta\right)=\cos\theta$;
for $n=5$ and $m=1$, $B_{5}^{1}\left(\theta\right)=3\cos^{2}\theta-1$;
for $n=6$ and $m=4$, $B_{6}^{1}\left(\theta\right)=1$, $B_{6}^{2}\left(\theta\right)=3\cos^{2}\theta-1$,
$B_{6}^{3}\left(\theta\right)=3\cos^{2}\theta+1$ and
$B_{6}^{4}\left(\theta\right)=9\cos^{2}\theta-1$; and,
for $n=7$ and $m=4$, $B_{7}^{1}\left(\theta\right)=\cos^{2}\theta$,
$B_{7}^{2}\left(\theta\right)=\cos^{2}\theta-1$, $B_{7}^{3}\left(\theta\right)=\cos^{3}\theta$
and $B_{7}^{4}\left(\theta\right)=3\cos\theta-2\cos^{3}\theta$ \cite{buckingham1967}. The errors in the long-range potential fitting for the Li + CaF and Ca + LiF arrangements are 2.55 and 1.86 cm$^{-1}$, respectively. In addition, the long-range and short-range potentials are connected smoothly with a switch function. Specifically, we have

\begin{equation}
V_{\mathrm{pes}}=s_{abc}V_{abc}+\left(1-s_{abc}\right)\left(V^{(1)}_{a}+V^{(2)}_{bc}+V_{\mathrm{LR}}\right),
\end{equation}

\noindent where the arrangement-dependent switching function, $s_{abc}$, is defined as

\begin{equation}
s_{abc}\left(r_{ac}\right)=\frac{1-\tanh\left[\gamma_{s}\left(r_{ac}-r_{s}\right)\right]}{2}.
\end{equation}

\noindent When $bc$ = CaF, $\gamma_{s}=1$ \textit{a.u.}$^{-1}$ and $r_{s}=18$ \textit{a.u.} are used in the interval $0^{\circ} \le \theta \le 45^{\circ}$; $\gamma_{s}=1$ \textit{a.u.}$^{-1}$ and $r_{s}=13$ \textit{a.u.} within $45^{\circ} < \theta \le 75^{\circ}$; $\gamma_{s}=2$ \textit{a.u.}$^{-1}$ and $r_{s}=11$ \textit{a.u.} in $75^{\circ} < \theta \le 180^{\circ}$. Likewise, for $bc$ = LiF, $\gamma_{s}=0.8$ \textit{a.u.}$^{-1}$ and $r_{s}=14$ \textit{a.u.} are used within $0^{\circ} \le \theta \le 180^{\circ}$.

\begin{figure}[H]
\begin{centering}
\includegraphics[scale=0.23]{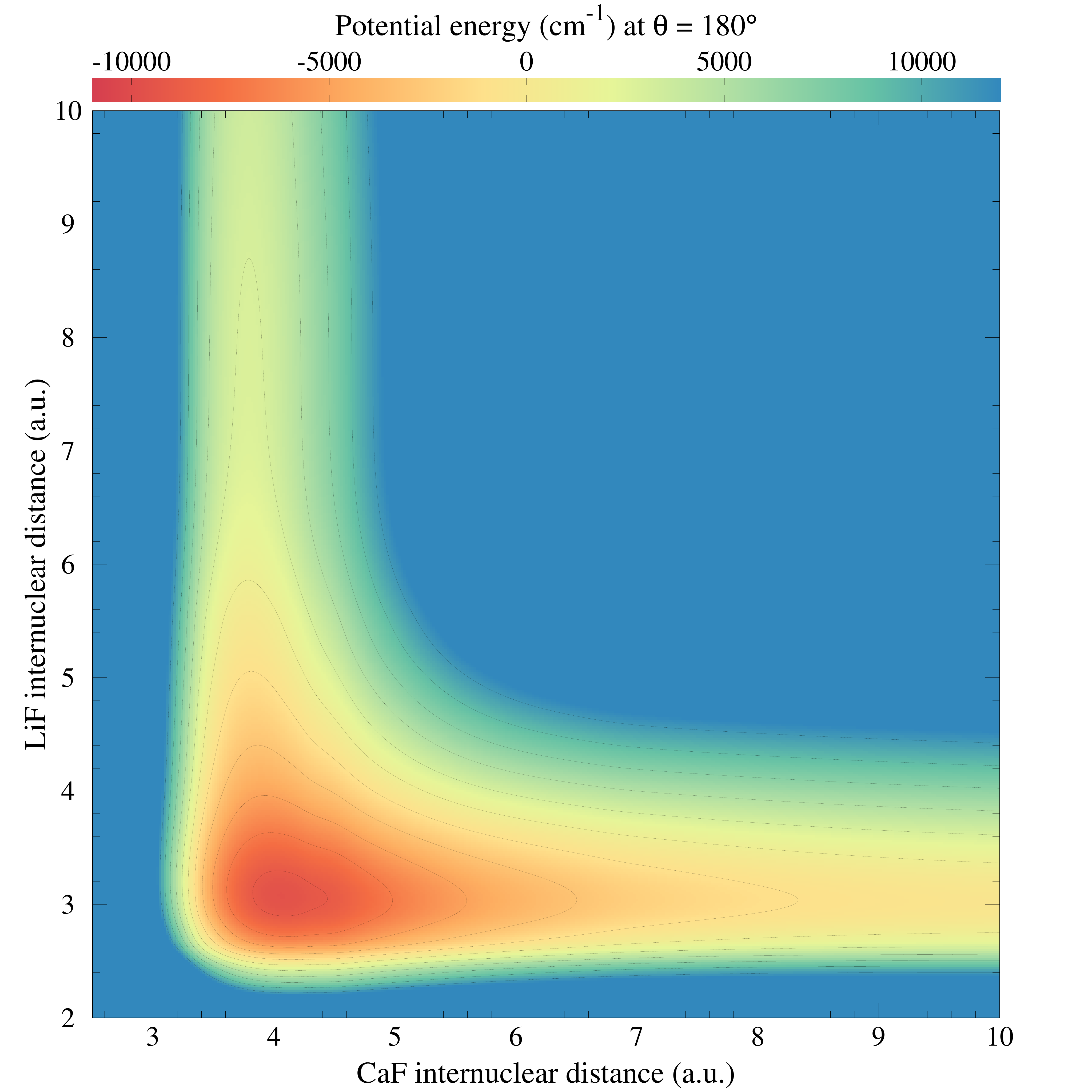}
\par\end{centering}
\caption{\label{fig:PESHeatmap180}Global PES, in cm$^{-1}$, as a function of the internuclear distances of LiF and CaF, in \textit{a.u.}, at the fixed angle $\theta=180^{\circ}$ centered at the F atom. Isolines varying every 2000 cm$^{-1}$ from -9000 cm$^{-1}$ (innermost) to 11000 cm$^{-1}$ (outermost).}
\end{figure}

\begin{figure}[H]
\begin{centering}
\includegraphics[scale=0.23]{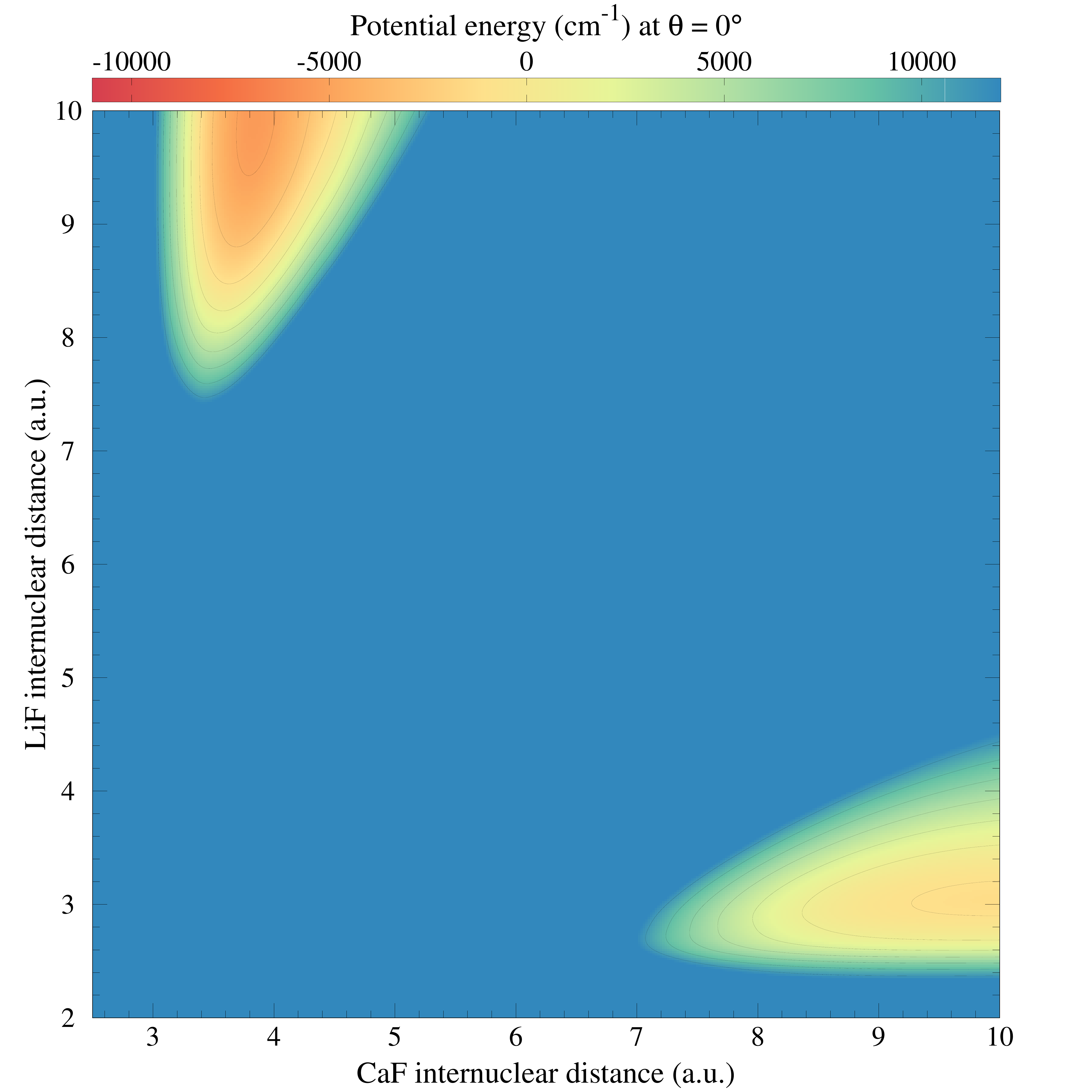}
\par\end{centering}
\caption{\label{fig:PESHeatmap0}The same as in Fig. (\ref{fig:PESHeatmap180}) but for $\theta=0^{\circ}$.}
\end{figure}

Figure (\ref{fig:PESHeatmap180}) depicts the contour plot of the PES, produced by the fitting procedure described above, as a function of the respective internuclear distances of LiF and CaF at the fixed bond-bond angle of $\theta=180^{\circ}$ centered at the F atom. The red-shaded regions are associated with the lowest (attractive) values of the potential whereas the blue areas correspond to higher energies. For the purpose of the scattering calculations presented below, in what follows, the zero-energy of the PES is shifted to correspond to the energy of LiF at the equilibrium position of $r_{\mathrm{LiF}}=$ 3.0 \textit{a.u.}, yellow area in Fig. (\ref{fig:PESHeatmap180}). The light-green region is for the CaF potential well with equilibrium position of $r_{\mathrm{CaF}}=$ 3.8 \textit{a.u.}, about 4040 cm$^{-1}$ above zero, excluding zero point energy (ZPE). The LiCaF potential well, where the three atoms are in close proximity, reaches a minimum value (about -10931.7 cm$^{-1}$) at slightly displaced diatomic distances, namely $r_{\mathrm{LiF}}=$ 3.18 and $r_{\mathrm{CaF}}=$ 4.06 \textit{a.u.}, in a near T-shape geometry at $\theta=104.5^{\circ}$ -- not shown but very similar to those contours of Fig. (\ref{fig:PESHeatmap180}). In addition, Fig. (\ref{fig:PESHeatmap0}) illustrate the strong anisotropic character of the system in which the PES is mostly repulsive for a collinear approach of the Li atom towards Ca $(\theta=0^{\circ})$ in contrast with the attractive character whenever approaching on the F side as shown in Fig. (\ref{fig:PESHeatmap180}).

\subsection{Adiabatically adjusting principal axis hyperspherical (APH) method}
\label{subsec:aph3d}

As mentioned above, the APH3D code is utilized  to model the
title reaction using the PES described in the previous section. A
somewhat detailed description of the numerical aspects and convergence
criteria is provided in the next section whereas, for the sake of
completeness, a brief overview on the implementation of APH3D is given
here. The description provided below is along the lines of
that given in our recent work on the H + D$_{2}$ chemical reaction
\cite{silva2022a}, however, an in-depth discussion of the hyperspherical
coupled-channel equations, as implemented on APH3D, has been given
by Kendrick and co-workers in many occasions \cite{pack1987,kendrick1999,kendrick2000,kendrick2001,kendrick2016a,kendrick2003a,kendrick2003b,kendrick2016b,kendrick2018a,kendrick2018b,kendrick2019}.

The LiCaF Hamiltonian is written in APH coordinates of Pack and Parker \cite{pack1987}.
The hyperradius, $\rho$, describing the radial atom-diatom relative
motion is partitioned into an inner region, using Smith-Johnson hyperspherical
coordinates \cite{smith1962b,kendrick1999}, where collision-induced
re-arrangement is more likely to occur. In the outer region, where the different atom-diatom arrangement channels are largely decoupled,
Delves hyperspherical coordinates \cite{delves1958,delves1960,parker2002} are employed. The six-dimensional three-body
problem is reduced to a set of coupled equations
along the scattering coordinate, $\rho$, with $\rho$ discretized
in a grid of $N$ sectors. The eigenvalues associated to the remaining
five internal degrees of freedom are used as the effective
set of coupled potentials driving the relative motion along $\rho$.

The 5D eigenvalue problem is solved in the APH region by means of
an implicitly restarted Lanczos method \cite{sorensen1992,maschhoff1996},
whereas the corresponding eigenvalues within the Delves region are
evaluated using a 1D Numerov propagator \cite{johnson1977}. Once
a sufficiently large set of coupled potentials are evaluated in both
regions for all sectors, as well as all sector-to-sector overlap matrices,
the resulting set of radial coupled equations is solved using Johnson's
log--derivative method \cite{johnson1973}, first from $\rho_{\mathrm{min}}$
to $\rho_{\mathrm{match}}$. At $\rho_{\mathrm{match}}$ the numerical
solutions from the outermost sector of the APH region are projected
onto solutions at the innermost sector of the Delves region. The propagation
is continued from $\rho_{\mathrm{match}}$ to $\rho_{\mathrm{max}}$,
a sufficiently large value of $\rho$ where the interaction potential
is negligible. At $\rho_{\mathrm{max}}$ all channels (from all arrangements)
are numerically decoupled, scattering boundary conditions are applied,
the log-derivative solutions are projected onto solutions associated
with each asymptotic diatomic state, written in ordinary Jacobi coordinates,
yielding a scattering matrix \cite{pack1987}. The procedure is repeated
independently for each value of the total angular momentum quantum
number $J$ and its parities, good quantum numbers in the absence
of external fields. However, as explained below, for the present work,
only the (even) $J=0$ case is addressed. Moreover, the basis sets
for both APH and Delves regions are independent of collision energy
and, therefore, evaluated only once.

Due to the fact that low-lying collisional channels, with relatively
higher kinetic energies, are associated with highly oscillatory components
of the scattering wavefunction, particular attention is given below
to the number of sectors, $N$, and the grid step size used, $\Delta\rho_{\mathrm{aph}}$
and $\Delta\rho_{\mathrm{delves}}$. In addition, combined with the
usual outward sector-to-sector integration of the Schr$\ddot{\mathrm{o}}$dinger
equation, an intra-sector subdivision of the grid is employed in both APH and Delves regions. Thus, for the $n^{\mathrm{th}}$ sector, of length $\Delta\rho$,
defined within the $\rho_{\mathrm{left}}^{n}$ and $\rho_{\mathrm{right}}^{n}$
boundaries, with $\rho_{\mathrm{left}}^{0}=\rho_{\mathrm{min}}$,
$\rho_{\mathrm{left}}^{n}=\rho_{\mathrm{right}}^{n-1}$ , $\rho_{\mathrm{right}}^{n}=\rho_{\mathrm{left}}^{n+1}$,
$\rho_{\mathrm{right}}^{N-1}=\rho_{\mathrm{max}}$ and $n=0,1,2,\ldots,N-1$,
the grid is further subdivided into $N_{\mathrm{steps}}$ per wavelength,
$\lambda_{\mathrm{max}}=\nicefrac{2\pi}{k_{\mathrm{max}}}$, where $k_{\mathrm{max}}$
is the maximum value of the wave vector considered, \textit{i.e.}

\begin{equation}
\frac{\hbar^{2}k_{\mathrm{max}}^{2}}{2\mu}=E_{\mathrm{max}}.\label{eq:MaxWavenumber}
\end{equation}

In Eq. (\ref{eq:MaxWavenumber}) $\mu$ is the atom-diatom reduced mass and $E_{\mathrm{max}}$
is a fixed parameter whose value is set as high as the asymptotic energy
of the highest closed channel included in the set of coupled equations,
such that all channels are well described. Therefore, in what follows,
we shall also determine the optimal value of $N_{\mathrm{steps}}$
such that the grid-dependent physical description of the problem remains
unaltered, \textit{i.e.} a proper description of the smallest periods
of oscillation of the wavefunction is included.

\subsection{Numerical considerations}

Despite a proper time-independent quantum formalism that takes into
consideration the doublet (or higher) spin multiplicity is available in
the domain of inelastic collisions \cite{alexander1982a,alexander1982b,corey1983,tscherbul2006,lopez-duran2008,hernandez-vera2017,zuo2020},
an implementation of the reactive counterpart of the problem is not. Therefore, we shall use the formalism for collisions between a $^{1}\Sigma^{+}$ molecule and a structureless atom. Such assumptions have been proven
valid in certain context for inelastic collisions \cite{hernandez-vera2017}.
In what follows we make a few considerations for the case-study at
hand.

Due to the null projection of the electronic orbital angular momentum
of CaF on its internuclear axis, $\Lambda(\Sigma^{+})=0$, and the
absence of a nearby electronic $^{2}\Pi$ state, only electrostatic
interactions are expected to play a significant role on the internal
structure of the molecule. Thus, the CaF effective (angular) Hamiltonian (neglecting vibrational and Stark terms)
could be approximated as \cite{frosch1952,radford1964}

\begin{equation}
\begin{aligned}H_{\mathrm{CaF}} \approx B_{e}N^{2} + \gamma\mathbf{\left(S\cdot N\right)} + b\mathbf{\left(S\cdot I\right)} + c\left(S_{z}I_{z}\right) + f\left(\mathbf{I\cdot N}\right),
\end{aligned}
\label{eq:CaFInternalHamiltonian}
\end{equation}
where $\mathbf{N}$ is the diatomic rotational angular momentum, $\mathbf{S}$
is the electronic spin angular momentum with $S_{z}=\nicefrac{1}{2}$ being the spin
component along a given $z$-axis parallel to the internuclear axis,
$\mathbf{I}$ is the nuclear spin with an $I_{z}=\nicefrac{1}{2}$
component (due to the $^{19}$F isotope), and $B_{e}$ is the diatomic rotation
constant. The parameters $\gamma$, $b$, $c$ and $f$ are the strength coefficients
for the electronic-spin-rotation, isotropic and anisotropic electronic-spin-nuclear-spin,
and nuclear-spin-rotation couplings, respectively. For convenience,
the strength coefficients for the $(\upsilon=0,N=0)$ manifold, as measured
by Childs and co-workers \cite{childs1981}, are given in Table (\ref{tab:CaFMolecularParams}).

\begin{table}[H]
\begin{centering}
\caption{\label{tab:CaFMolecularParams}Experimental values, in cm$^{-1}$, of the molecular parameters of CaF for the $\upsilon=0,N=0$ rovibrational manifold \cite{childs1981}.}
\begin{tabular}{rr}
 & \tabularnewline
\hline
\hline
 & \tabularnewline
 & $X^{2}\Sigma^{+},\upsilon=0,N=0$\tabularnewline
 & \tabularnewline
\cline{2-2}
 & \tabularnewline
$B_{e}$ & 0.343704\tabularnewline
$\gamma$ & 1.323$\times10^{-3}$\tabularnewline
$b$ & 3.642$\times10^{-3}$\tabularnewline
$c$ & 1.338$\times10^{-3}$\tabularnewline
$f$ & 9.593$\times10^{-7}$\tabularnewline
 & \tabularnewline
\hline
\end{tabular}
\par\end{centering}
\end{table}

As expected, the diatomic rotation constant is much larger than the
remaining coupling parameters (about 257 times larger than $\gamma$
and $c$, 94 times larger than $b$) and, therefore, the dominant
term. As a consequence, the $N=1$ rotational structure, within the
$\upsilon=0$ manifold, is predicted to lie about $2B_{e}\approx0.69$
cm$^{-1}$ (or 900 mK, neglecting higher-order centrifugal distortion
contributions) above the $N=0$ structure. Using either a Hund's case
(a) \cite{alexander1982a} or (b) \cite{corey1983} notation, both
$\mathbf{N}$ and $\mathbf{S}$ are generally well accepted to be
weakly coupled. As a consequence, collision-induced changes in either
the magnitude or the direction of the electronic spin, $\mathbf{S}$,
are unlikely to happen. However, a collision may induce sudden changes
in $\mathbf{N}$ and, due to the subsequent recoupling between $\mathbf{N}$
and $\mathbf{S}$, changes between the resultant parallel ($e$ parity)
and anti-parallel ($f$ parity) coupling schemes may occur \cite{alexander1982a}.
Since we are not properly describing the diatomic rotational structure,
we shall address collision energies well below the 900 mK threshold,
such that the $N=1$ rotational state will remain as a closed
channel.

In regard to the $N=0$ fine/hyperfine structure, for $\Lambda=0,I_{z}=S_{z}=\nicefrac{1}{2}$,
the predicted sublevels of $H_{\mathrm{CaF}}$ are associated to $j=\nicefrac{1}{2}$
and $F=0$ or 1 quantum numbers, where $\mathbf{j}=\mathbf{N}+\mathbf{S}$
(fine structure) and $\mathbf{F}=\mathbf{j}+\mathbf{I}$ (hyperfine
structure) \cite{childs1981}. However, as the collisions treated below are
explored in the absence of external fields and, given the equally
small electronic-spin-nuclear-spin interactions, $b$ and $c$, alongside
the negligible nuclear-spin-rotation interaction, $f$, the multiplet structure of the entrance channel is not considered henceforth. Even
if external fields were taken into account, a collision-induced Zeeman
relaxation of the $N=0$ rotational structure of CaF is expected to
vanish at first-order, being mostly a second-order process \cite{krems2004a}
and, therefore, it appears reasonable to neglect. However, it is worthwhile to note that, as neither the doublet spin
multiplicity of CaF nor that of Li is included, a resultant magnetic dipole-dipole interaction is also disregarded.
As such interactions are more prevalent in the ultracold regime of kinetic energies, our assumption of a pseudo $^{1}S+{}^{1}\Sigma$ colliding system implies a lower limit on the range of collision energies that can be studied here without compromises. As we shall see below, 1 mK is the minimal energy treated in this work.

Thus, from now on, we drop the use of the typical Hund's case (b) labeling of $N_{j}$
quantum numbers for the diatomic rotational structure in favor of
the $j$ rotational level (an integer, $j=0,1,\ldots$), as used in
the literature of singlet molecules.

Another aspect that we shall not describe in this work is higher values
of the total angular momentum, \textit{i.e.} $J>0$. The amount of
computational resources required to perform a $J=0$ calculation is
already substantial and the inclusion of higher $J$ values would
increase it  drastically, as we would be now required to handle
both even and odd parities of each non-zero $J$ case. Despite the
relatively low collision energies intended, yet well above the $s$-wave
regime, it is likely that a few $J$ values are still required
to secure convergence. As a consequence, the $J=0$ calculation
presented below may not be suited for a direct quantitative comparison
with experimental results. However, it is worthwhile to stress that
$J=0$ calculations have been proven to provide an insightful and
accurate qualitative description of collisional problems in the past
besides providing also the foundation for the optimization of certain
key numerical parameters.

\begin{figure}[H]
\begin{centering}
\includegraphics[scale=0.4]{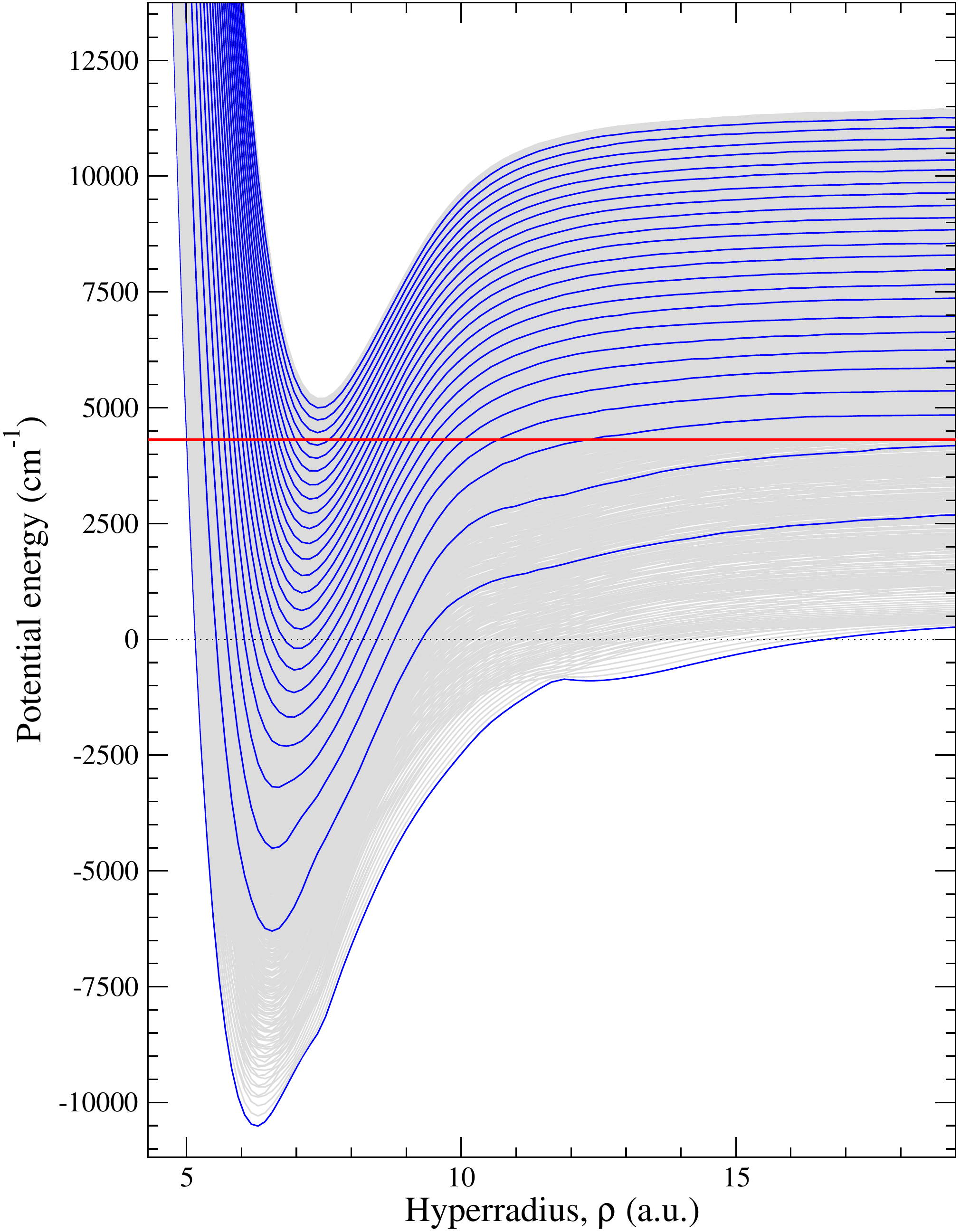}
\par\end{centering}
\caption{\label{fig:AphCoupledPotential}Set of the lowest 2500 APH coupled potential curves (cm$^{-1}$), in which every 100 curves are highlighted in blue, as functions of the hyperradius, $\rho$ ($a.u.$). The bold red line tags the CaF($\upsilon=0,j=0$) diatomic rovibrational level as a pseudo-$^{1}\Sigma^{+}$ molecule.}
\end{figure}

Thus, in the context of a pseudo-$^{1}\Sigma^{+}$ CaF molecule colliding
with a structureless Li atom for $J=0$, we then proceed to determine
the optimal parameters in order to describe the scattering wavefunction
in both APH and Delves regions. Figure (\ref{fig:AphCoupledPotential})
shows the final set of 2500 coupled potentials used in the APH region,
evaluated between $\rho_{\mathrm{min}}=4.5\:a.u.$ and $\rho_{\mathrm{match}}\approx19.9\:a.u.$
with 76 sectors varying logarithmically with a step size of $\Delta\rho_{\mathrm{aph}}=0.02$ atomic units.
For easy visualization every 100th level is shown in blue.
The bold red line represents the $(\upsilon=0$, $j=0)$ entrance
channel of CaF. Despite many high-lying  blue curves being asymptotically
correlated with internal states a few thousand wavenumber above the
entrance channel, they are strongly interacting at short range, $\rho\approx$
6-7 $a.u.$, and their inclusion is needed to achieve converged results. However, the inclusion of more channels increases
the time-complexity of solving the Schr$\ddot{\mathrm{o}}$dinger
equation by a few factors of $O\left(N_{\mathrm{channels}}^{3}\right)$,
where $N_{\mathrm{channels}}$ is the number of channels to be included
in the basis set of the scattering wavefunction. Thus, considering
the number of calculations required to probe other convergence aspects
(as shown below) and energy-dependent calculations, increasing the
number of channels would quickly become impractical.
If higher collision energies than those addressed here are of
interest, it is desirable to include more channels, mainly in the
range between $\rho_{\mathrm{min}}$ and $\rho\approx13.5\:a.u.$,
then projecting the solutions into a smaller basis set, as the one
used here, and resuming the propagation toward large distances. The
choice of $\Delta\rho_{\mathrm{aph}}=0.02\:a.u.$ will remain fixed
in the remainder of this work. This is mostly due the inherent higher computational
overhead of optimizing it as evaluation of different sets of coupled
potentials would be required. However, it is worth noting
that similar values have been used for converged calculations of systems
as heavy as the one treated here and for somewhat similar grid parameters,
\textit{e.g.} $\Delta\rho_{\mathrm{aph}}=0.01\:a.u.$ used for Rb +
K$_{2}$ by Croft \textit{et al.} \cite{croft2017} and $\Delta\rho_{\mathrm{aph}}=0.012\:a.u.$
used for Li + LiNa by Kendrick \textit{et al.} \cite{kendrick2021}
(logarithmic scales used in all cases).

Beyond $\rho_{\mathrm{match}}$, now within the Delves region, the
additional concern of how the scattering characteristics may vary
with respect to $\rho{}_{\mathrm{max}}$, due to the lower collision
energies, should be addressed. Fortunately, by inspecting Fig. (\ref{fig:AphCoupledPotential})
again, the set of coupled potential curves is seen to present a somewhat
parallel behavior with respect to one another, at $\rho_{\mathrm{match}}$,
mostly due to the smaller couplings, as $\rho\rightarrow\rho_{\mathrm{max}}$.
This aspect in particular suggests that a much smaller number of channels
may be included in the basis set of the Delves region. In the present
work, 600 channels are utilized to solve the Schr$\ddot{\mathrm{o}}$dinger
equation along the Delves part, of which 396 are closed channels.
Possibly, the number of basis functions used in the Delves region
exceeds the requirements to obtain well converged numerical results
at $J=0$ and can be utilized also to describe collisions at higher
energies.

Asymptotically, at $\rho_{\mathrm{max}}$, the diatomic eigenstates
used as basis set comprises up to $\upsilon_{\mathrm{max}}=7$ and
$j_{\mathrm{max}}=68\:\left(\upsilon=0\right)$ for LiF, from which
$\left(\upsilon=0,j=55\right)$, $\left(\upsilon=3,j=33\right)$ and
$\left(\upsilon=4,j=21\right)$ are the highest open rovibrational
manifolds. Whereas for CaF, $\upsilon_{\mathrm{max}}=3$ and $j_{\mathrm{max}}=77\:\left(\upsilon=0\right)$
are utilized with all but the entrance channel energetically closed.
This smaller basis set, when compared with the one
used in the APH region, allows us to explore the convergence criteria with respect to $N_{\mathrm{steps}}$
(APH and Delves regions) and $\rho_{\mathrm{max}}$ (Delves region
only) in great detail as described below.

\begin{figure}[H]
\begin{centering}
\includegraphics[scale=0.4]{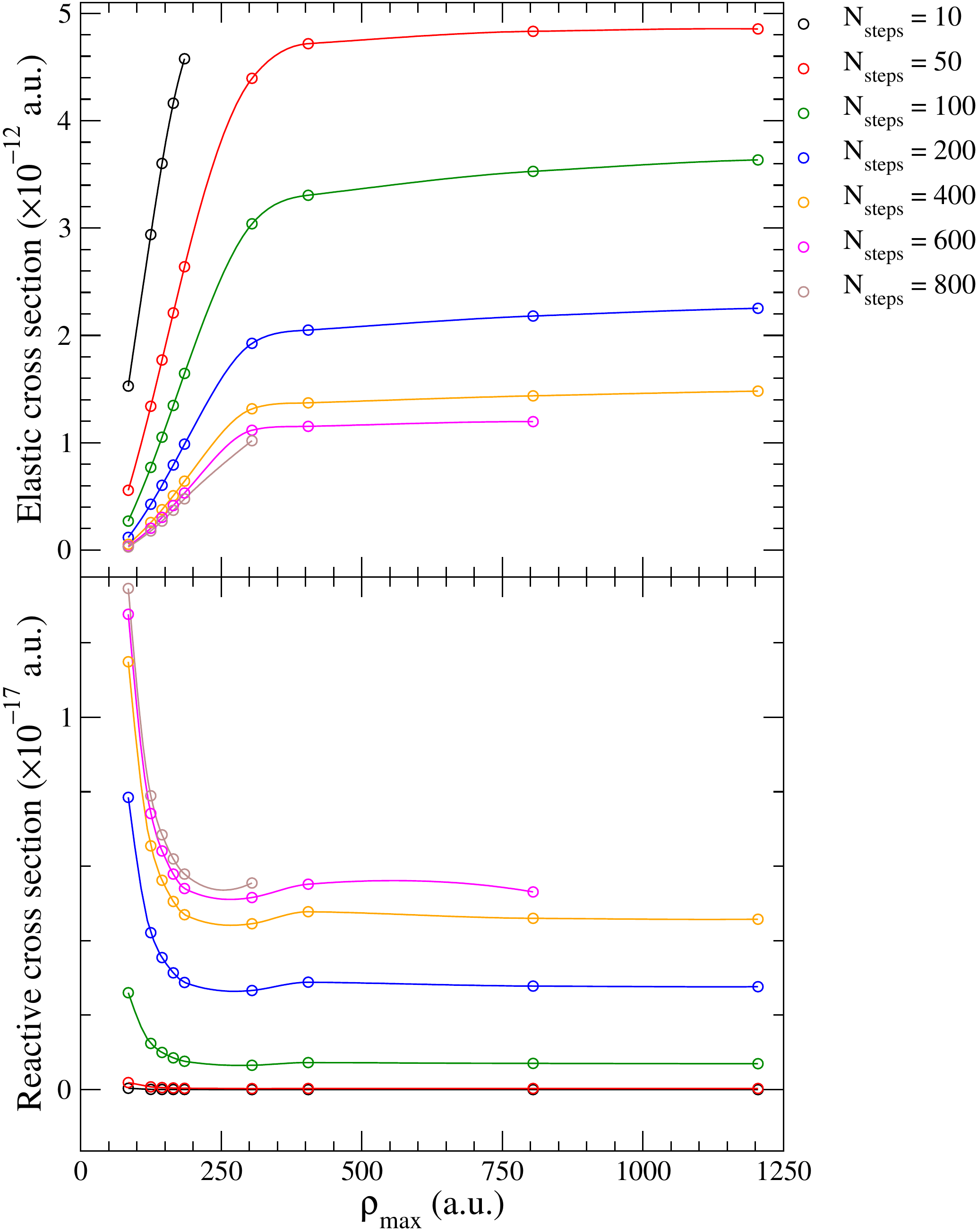}
\par\end{centering}
\caption{\label{fig:StepsVsRhomaxConvergence}Upper panel: Elastic component of the cross section (\textit{a.u.}) for the Li + CaF collisions at 1 mK as a function of $\rho_{\mathrm{max}}$ (\textit{a.u.}) for $N_{\mathrm{steps}}$ = 10 (black curve), 50 (red curve), 100 (green curve), 200 (blue curve), 400 (orange curve), 600 (magenta curve) and 800 (brown curve). Lower panel: Reactive component of the cross section (\textit{a.u.}) for the Li + CaF$\left(\upsilon=0,j=0\right)$ $\protect\longrightarrow$ \textcolor{black}{Ca + LiF}$\left(\upsilon^{\prime}=0,j^{\prime}=0\right)$ chemical reaction at 1 mK as a function of $\rho_{\mathrm{max}}$ (\textit{a.u.}) using the same color code as in the upper panel. Circles are raw CC calculations, curves are Akima splines to enhance visualization.}
\end{figure}

We have used a reference collision energy of 1 mK relative to the $\left(\upsilon=0,j=0\right)$
entrance channel of CaF to compute both elastic and reactive cross
sections for sets of $N_{\mathrm{steps}}=$ 10, 50, 100, 200, 400,
600 and 800 combined with $\rho_{\mathrm{max}}\approx$ 85, 125, 155,
185, 305, 405, 805 and 1250 atomic units. We note that, all parameters except $N_{\mathrm{steps}}$
are held fixed in the APH region, whereas in the Delves region, each
$N_{\mathrm{steps}}$ choice is combined with an increasing number of
sectors varying linearly at fixed steps of $\Delta\rho_{\mathrm{delves}}=0.2$ atomic units.
The result is presented in Fig. (\ref{fig:StepsVsRhomaxConvergence}),
where each curve corresponds to a given value of $N_{\mathrm{steps}}$
and the cross section is plotted as a function of $\rho_{\mathrm{max}}$.
An inspection of the upper panel (elastic component) suggests a somewhat
strong dependence on both parameters, as expected, and the cross section
converges from below to its optimal value between $\rho_{\mathrm{max}}\approx250$
and 350 \textit{a.u.} with values of $N_{\mathrm{steps}}>600$
yielding comparable results. The reactive cross sections for
the LiF$\left(\upsilon^{\prime}=0,j^{\prime}=0\right)$ exit channel are presented in the lower panel of Fig. (\ref{fig:StepsVsRhomaxConvergence})
where a similar convergence pattern (now from above) is evident, except
that the set of calculations with $N_{\mathrm{steps}}=10$ and 50 are completely
unable to describe the reaction. Thus, $N_{\mathrm{steps}}>200$ and,
ideally, 600 is recommended.

\begin{figure}[H]
\begin{centering}
\includegraphics[scale=0.4]{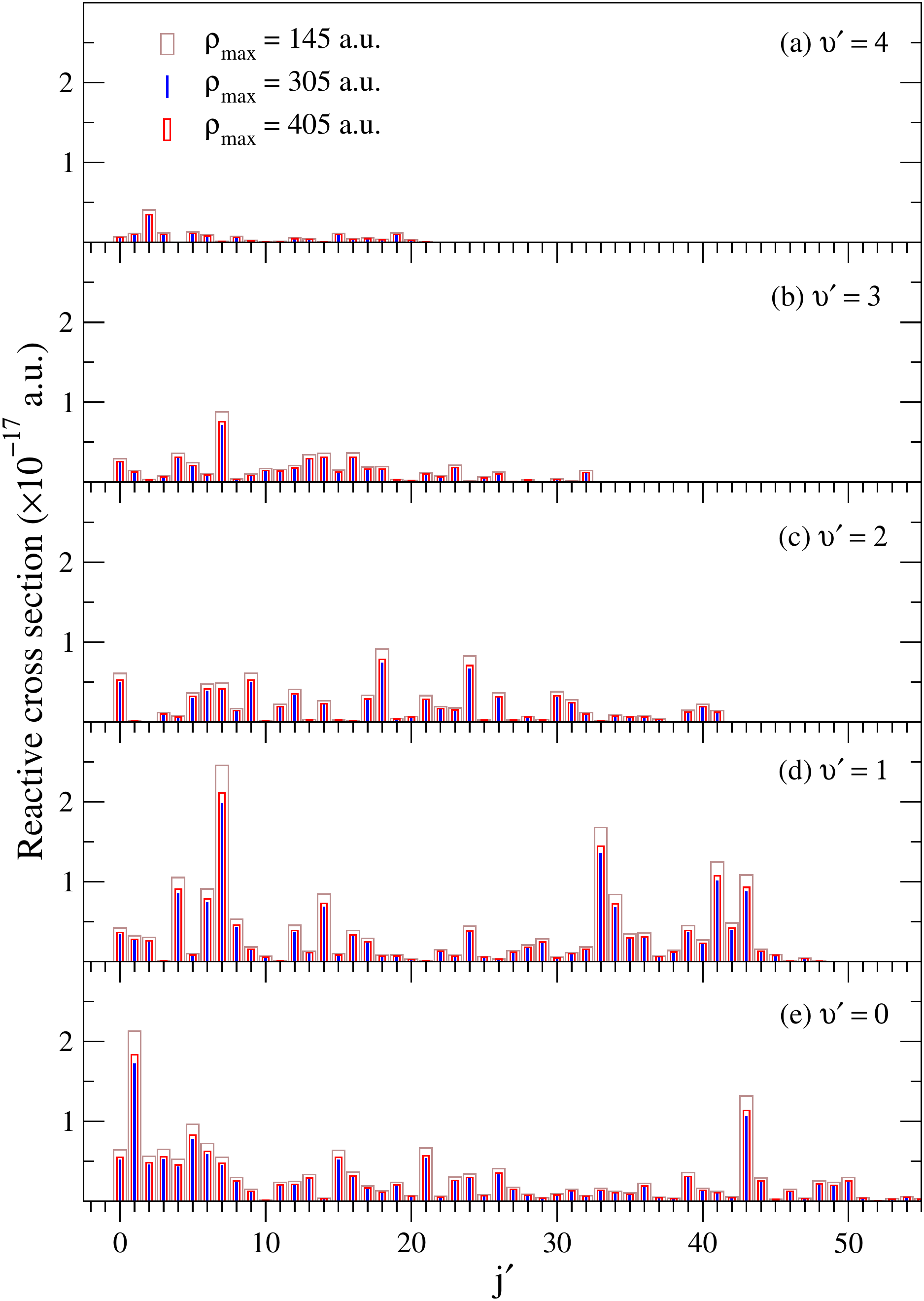}
\par\end{centering}
\caption{\label{fig:OneMillikelvinAngularDistribution}Reactive cross sections for the Li + CaF$\left(\upsilon=0,j=0\right)\protect\longrightarrow$ \textcolor{black}{Ca + LiF}$\left(\upsilon^{\prime},j^{\prime}\right)$ collision at 1 mK $\left(N_{\mathrm{steps}}=600\right)$ as functions of $j^{\prime}$. Brown bars are used for $\rho_{\mathrm{max}}\approx145$ \textit{a.u.}, blue bars for $\rho_{\mathrm{max}}\approx305$ \textit{a.u.} and red bars for $\rho_{\mathrm{max}}\approx405$ \textit{a.u.} with panels (a)-(e) corresponding to $\upsilon^{\prime}=$ 4, 3, 2, 1 and 0, respectively.}
\end{figure}

For the sake of simplicity, the convergence behavior of cross sections
for other choices of $\upsilon^{\prime}$ and $j^{\prime}$ of LiF
are not shown but they possess virtually identical patterns as those
observed in the lower panel of Fig. (\ref{fig:StepsVsRhomaxConvergence}).
Instead, in Fig. (\ref{fig:OneMillikelvinAngularDistribution}), the
reactive cross section for all open $\upsilon^{\prime}$ (panels)
and $j^{\prime}$ (abscissa) exit channels at the fixed values of
$\rho_{\mathrm{max}}=145,305,405$ \textit{a.u.} and $N_{\mathrm{steps}}=600$
is presented. A qualitative description of Fig. (\ref{fig:OneMillikelvinAngularDistribution})
is given in the next section whereas, for now, suffice to observe
that each independent calculation (blue, red and brown) captures
virtually identical branching ratios over both $\upsilon^{\prime}$
and $j^{\prime}$ implying that the calculations are numerically
stable in order to infer the actual physical aspects of these collisions.

\section{Results and Discussion}
\label{sec:results}

\begin{figure}[H]
\begin{centering}
\includegraphics[scale=0.4]{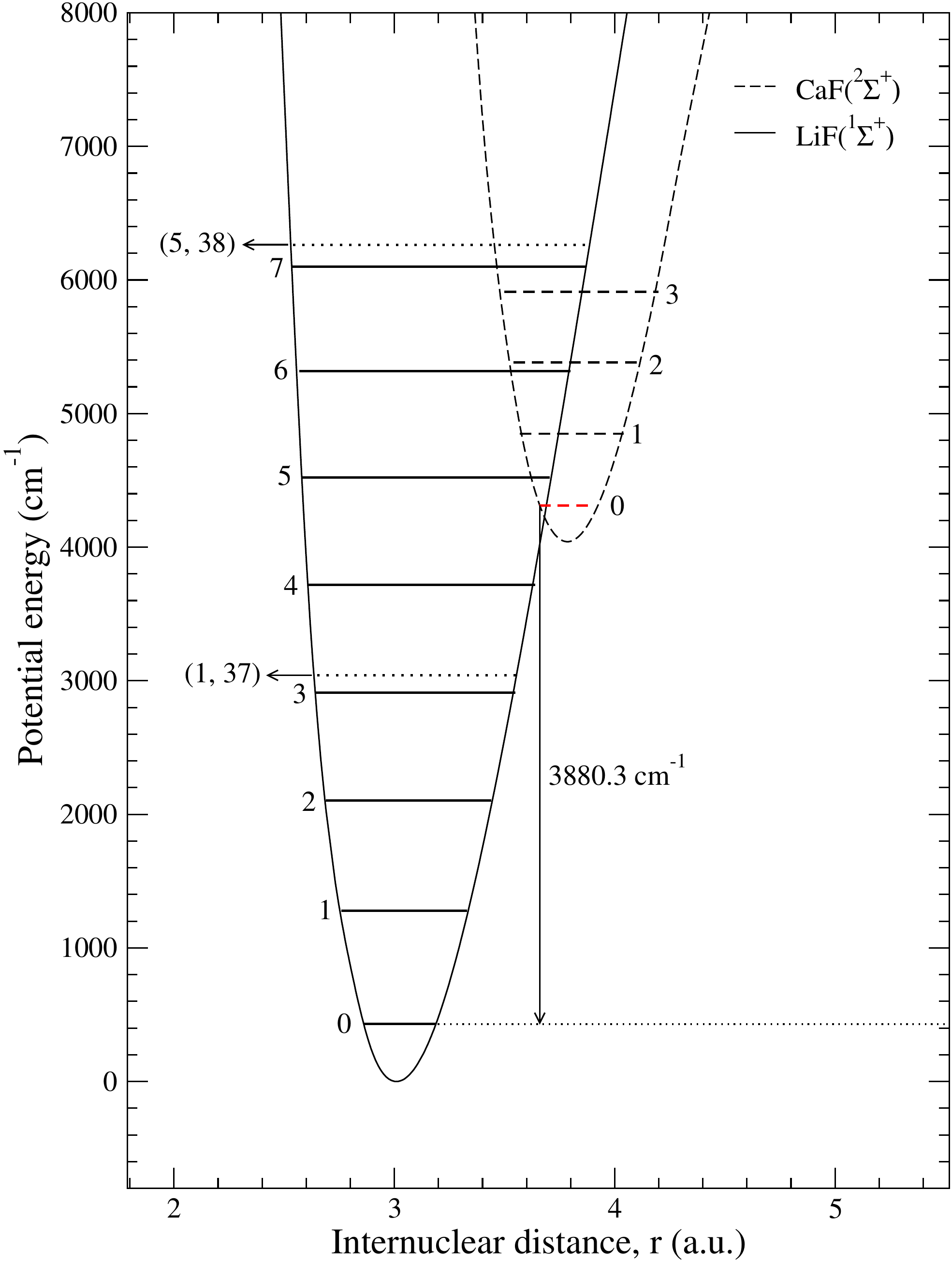}
\par\end{centering}
\caption{\label{fig:PECs} Diatomic potential energy curves for CaF$(X^{2}\Sigma^{+})$ (black dashed curve) and LiF$(X^{1}\Sigma^{+})$ (black solid curve), both in cm$^{-1}$, as functions of $r$ with the respective third atoms (Li and Ca) at a distance of 100 \textit{a.u.} $(\theta=0^\circ)$. Horizontal bold (solid and dashed) lines tag the diatomic vibrational levels $(j=0)$ for the respective quantum number displayed. Horizontal bold dotted lines tag the vibrational levels corresponding to rotational states $j=37$ $(\upsilon=1)$ and $j=38$ $(\upsilon=5)$.}
\end{figure}

First, we address how well the PES can reproduce the asymptotic PECs for the CaF$(X^{2}\Sigma^{+})$ and LiF$(X^{1}\Sigma^{+})$ subsystems at large atom-diatom separations. The bottom of each PEC is presented in Fig. (\ref{fig:PECs}), in which case the global dissociation limit corresponding to Li$(^{2}S)$ + Ca$(^{1}S)$ + F$(^{2}P)$, $\approx47986$ cm$^{-1}$, is not shown. Due to our earlier choice of using the LiF energy at the equilibrium position as the zero-energy in the scattering calculations, the $47986$ cm$^{-1}$ limit also corresponds to the relative dissociation energy of LiF, $D_{e}$. For comparison purposes a list of a few selected values of equilibrium positions and dissociation energies, for both LiF and CaF electronic ground states, are collected in Table (\ref{tab:PECComparison}). In the particular case of CaF, electronic structure data is somewhat scarce and/or dated. Yet, by inspection of Table (\ref{tab:PECComparison}), we do observe a reasonably good agreement between our calculations and those from literature, in particular, the recent results of Sardar and co-workers \cite{sardar2022}. As expected, the dissociation energies appear to vary more broadly, within $\approx2000$ cm$^{-1}$ among the various studies, with our result well within that range. Despite that the error appears to be relatively small if the actual total depth of the potential is taken into consideration ($\approx47986$ cm$^{-1}$ for LiF), the data collected in Table (\ref{tab:PECComparison}) seem to suggest that it is relatively harder to properly reproduce the LiF well depth than that of CaF. As investigated in great detail by Varandas \cite{varandas2009}, the LiF electronic ground state is not trivial, manifesting a predominant ionic character at the equilibrium position and avoid crossing the $2^{1}\Sigma^{+}$ excited state (whose nature is essentially covalent) at relatively short ranges ($r_{\mathrm{LiF}}\approx14$ atomic units). Moreover, it is asymptotically correlated with an additional $^{1}\Pi$ state.

Using the PECs shown in Fig. (\ref{fig:PECs}), an energy splitting of about 0.65 cm$^{-1}$ between the first two rotational levels of CaF is predicted, in the $\upsilon=0$ vibrational manifold, which suggests an effective diatomic rotational constant of $B_{e}=0.65/2=0.325$ cm$^{-1}$ and, thus, is within 0.02 cm$^{-1}$ from the value measured by Childs \textit{et al.} -- see Table (\ref{tab:CaFMolecularParams}). Similarly, an effective diatomic rotational constant of $B_{e}=1.29$ cm$^{-1}$ is predicted for LiF, which agrees reasonably well with the measured value of 1.3452576 cm$^{-1}$ \cite{huber1979}, but overall these evidence seems to suggest that the shape of the PECs (and their bound states) is equally satisfactory. The energy levels for the vibrational states utilized in the scattering calculations are tagged with horizontal lines in Fig. (\ref{fig:PECs}) and, for the sake of clarity, only $j=0$ cases are displayed, except for $j=37$ ($\upsilon=1$) and $j=38$ ($\upsilon=5$, the highest basis function taken into account). By including the respective ZPEs of each molecule a total exothermicity of about 3880.3 cm$^{-1}$ is expected and, therefore, it is a few hundred wavenumbers below the earlier prediction of 4440 cm$^{-1}$ by Kosicki and co-workers \cite{kosicki2017}.

\begin{table*}
\begin{centering}
\caption{\label{tab:PECComparison}Equilibrium positions ($r_{\mathrm{LiF}}$ and $r_{\mathrm{CaF}}$, in \textit{a.u.}) and dissociation limits ($D_{e}$, in cm$^{-1}$) along with the method utilized (see the respective references for details). The original values, in $\mathring{\mathrm{A}}$ and eV, are given between parentheses with the factors 1\textit{ a.u.} = 1 a$_{0}$ = 0.52917721092 $\mathring{\mathrm{A}}$ and 1\textit{ a.u.} = 27.211385 eV = 219474.63137054 cm$^{-1}$ applied. $^{a}$The original $D_{e}$ value is converted from kcal/mol (135.8) and the geometry optimization is made at the MP2(full)/6311+G{*} level. $^{b}$Within $\pm$ 0.3 eV ($\pm$ 2419.7 cm$^{-1}$).}
\begin{tabular}{rrrrrrrrrrr}
 &  &  &  &  &  &  &  &  &  & \tabularnewline
\hline
\hline
 &  &  &  &  &  &  &  &  &  & \tabularnewline
 &  &  &  &  &  &  & LiF &  &  & CaF \tabularnewline
 &  &  &  &  &  &  &  &  &  & \tabularnewline
\cline{6-11}
 &  &  &  &  &  &  &  &  &  & \tabularnewline
 & Ref. &  & Method &  & $r_{\mathrm{LiF}}$ &  & $D_{e}$  &  & $r_{\mathrm{CaF}}$ & $D_{e}$ \tabularnewline
 &  &  &  &  &  &  &  &  &  & \tabularnewline
\cline{2-11}
 &  &  &  &  &  &  &  &  &  & \tabularnewline
 & This work &  & MRCI+Q/CASSCF &  & 3.0 &  & 47985.4 &  & 3.8 & 43944.6\tabularnewline
 &  &  &  &  &  &  &  &  &  & \tabularnewline
 & \cite{sardar2022} &  & MRCI+Q/CASSCF &  & &  & &  & 3.7 & 43672.0\tabularnewline
 &  &  &  &  &  &  &  &  &  & \tabularnewline
 & \cite{hou2018} &  & Semi-empirical &  & &  & &  & 3.69 & 44203.5\tabularnewline
 &  &  &  &  &  &  &  &  &  & \tabularnewline
 & \cite{sardar2022} &  & Empirical &  & &  & & & 3.71 &\tabularnewline
 &  &  &  &  &  &  &  &  &  & \tabularnewline
 & \cite{rao1983} &  & Empirical &  &  &  &  &  &  & 44111.3\tabularnewline
 &  &  &  &  &  &  &  &  &  & \tabularnewline
 & \cite{yang2007} &  & MRCI, CBS &  &  &  &  &  & 3.78 (2.0005) & \tabularnewline
 &  &  &  &  &  &  &  &  &  & \tabularnewline
 & \cite{yang2004} &  & B3LYP/BS3, HP &  &  &  &  &  & 3.68 (1.9485) & 45752.6 (5.6726)\tabularnewline
 &  &  &  &  &  &  &  &  &  & \tabularnewline
 & \cite{partridge1984} &  & HF/STO &  & 2.929 (1.5500) &  & 49200.0 (6.1000) &  & 3.74 (1.9800) & 43957.0 (5.4500)\tabularnewline
 &  &  &  &  &  &  &  &  &  & \tabularnewline
 & \cite{varandas2009} &  & CAS-A7/XZ, CBS &  & 2.983 (1.5788) &  & 42940.1 (5.3239) &  &  & \tabularnewline
 &  &  &  &  &  &  &  &  &  & \tabularnewline
 & \cite{varandas2009} &  & MRCI-C3$_{2}$/XZ, CBS &  & 2.985 (1.5795) &  & 47823.8 (5.9294) &  &  & \tabularnewline
 &  &  &  &  &  &  &  &  &  & \tabularnewline
 & \cite{varandas2009} &  & MRCI-C3$_{0}$/cXZ, CBS &  & 2.933 (1.5524) &  & 49003.8 (6.0757) &  &  & \tabularnewline
 &  &  &  &  &  &  &  &  &  & \tabularnewline
 & \cite{varandas2009} &  & MRCI-C0/cXZ, CBS &  & 2.952 (1.5622) &  & 48953.8 (6.0695) &  &  & \tabularnewline
 &  &  &  &  &  &  &  &  &  & \tabularnewline
 & \cite{boldyrev1993} &  & PMP4/6-311+G$\left(2df\right)$ &  & 3.014 (1.5950) &  & 47496.7 (5.8888)$^{a}$ &  &  & \tabularnewline
 &  &  &  &  &  &  &  &  &  & \tabularnewline
 & \cite{varandas2009} &  & Empirical &  & 2.955 (1.5638) &  & 48393.3 (6.0000)$^{b}$ &  &  & \tabularnewline
 &  &  &  &  &  &  &  &  &  & \tabularnewline
 & \cite{varandas2009} &  & Empirical &  &  &  & 48554.6 (6.0200) &  &  & \tabularnewline
 &  &  &  &  &  &  &  &  &  & \tabularnewline
\hline
\end{tabular}
\par\end{centering}
\end{table*}

\begin{figure}[H]
\begin{centering}
\includegraphics[scale=0.4]{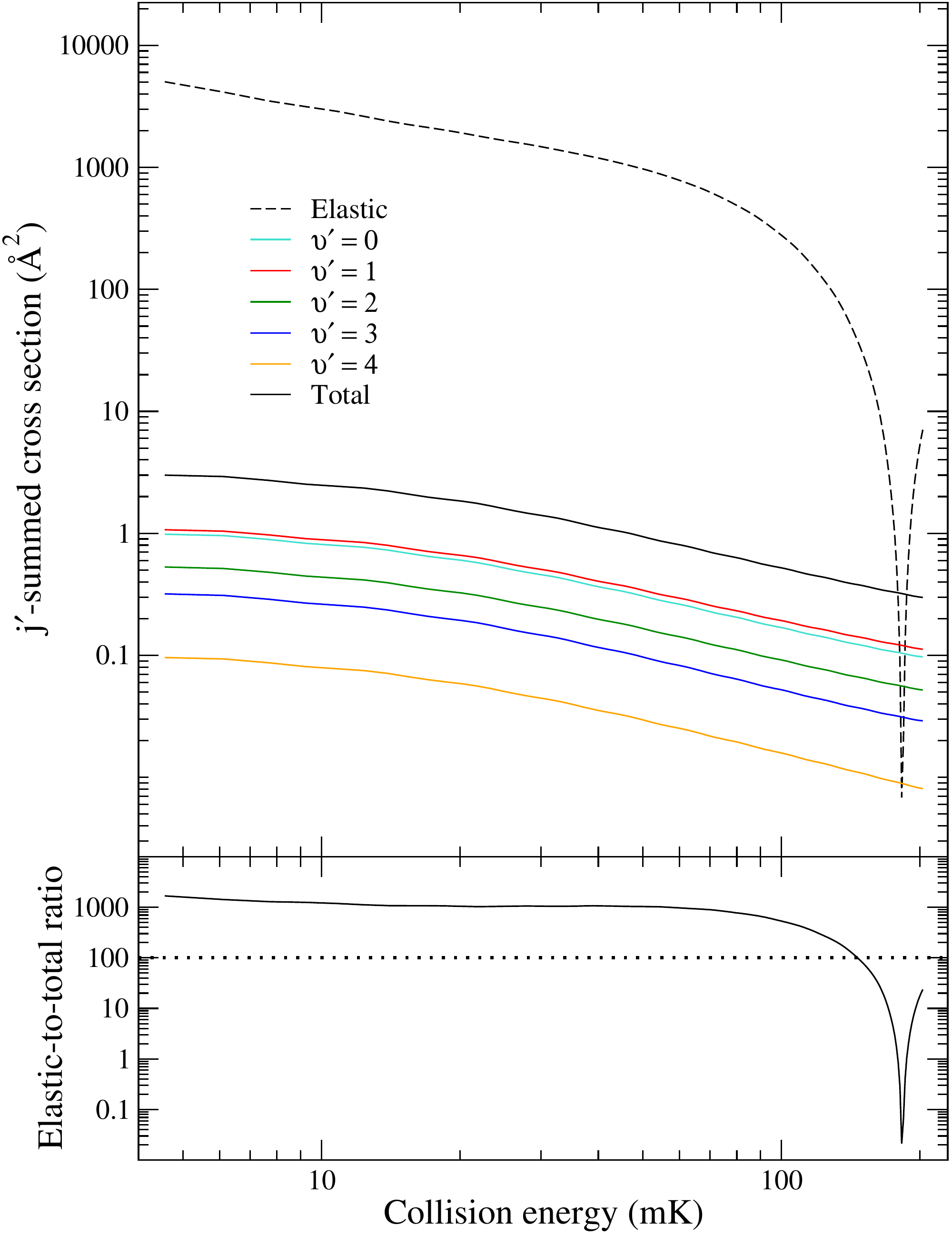}
\par\end{centering}
\caption{\label{fig:CrossSectionVsEnergyVsRatio}Upper panel: Reactive cross
section, in $\mathring{\mathrm{A}}^{2}$, for the Li + CaF$\left(\upsilon=0,j=0\right)\protect\longrightarrow$
\textcolor{black}{Ca + LiF}$\left(\upsilon^{\prime}\right)$ chemical
reaction, summed over $j^{\prime}$, as a function of the collision energy,
in mK; where, $J=0$, $N_{\mathrm{steps}}=600$, $\rho_{\mathrm{max}}=305$ \textit{a.u.},
$\upsilon^{\prime}=0$ (cyan curve), $\upsilon^{\prime}=1$ (red
curve), $\upsilon^{\prime}=2$ (green curve), $\upsilon^{\prime}=3$
(blue curve), $\upsilon^{\prime}=4$ (orange curve), the total summed
over $\upsilon^{\prime}$ (black curve) and elastic component (dashed
black curve). Lower panel: The elastic-to-total reactive cross section ratio as a function of the
collision energy.}
\end{figure}

In the remainder of the paper we describe the Li + CaF $\longrightarrow$
Ca + LiF chemical reaction with those parameters
described in the previous section. We will perform a scan on collision energy from 1 mK to 200
mK for the $\left(\upsilon=0,j=0\right)$ entrance channel of CaF,
in a grid of 128 points varying linearly, using $N_{\mathrm{steps}}=600$
and $\rho_{\mathrm{max}}=305$ \textit{a.u.}; the result of which is presented
in Fig. (\ref{fig:CrossSectionVsEnergyVsRatio}). The choice of the ground rovibrational state of CaF as the entrance channel for these collisions, at sufficiently small collision energies, rule out the occurrence of inelastic processes, such that the only non-elastic pathway is the chemical reaction. In addition, as the PES described above does not take into account the nearby $^{3}A^{\prime}$ electronic state (degenerate asymptotically), the influence of singlet-triplet nonadiabatic transitions and/or spin-exchange effects on the reaction presented below, if any, is disregarded. As no actual comparison with a measurement and/or other calculations is possible for now, those results presented below are not scaled by the typical $\nicefrac{1}{4}$ statistical weight factor of singlet entrance channels with respect to their triplet counterparts. That implies a hypothetical scenario in which 100\% of the colliding partners are prepared in the electronic ground state of the complex. In an actual experimental scenario, with no control of the initial spin, it is expected that up to 75\% of the collisions would undergo elastic and inelastic processess along the triplet PES whereas 25\% would undergo the reactive process described here. In the discussion presented below, the absolute value of the cross section is less relevant and we shall address aspects of the relative quantities such as (branching) ratios.

In the upper panel of Fig. (\ref{fig:CrossSectionVsEnergyVsRatio}),
the energy-dependence of the cross sections, summed over $j^{\prime}$,
is presented for each open manifold associated to $\upsilon^{\prime}=0$-4
of the LiF product (cyan, red, green, blue and orange curves),
whereas the total, summed over $j^{\prime}$ and $\upsilon^{\prime}$,
is denoted by the solid black curve. For comparison purposes the elastic
component of the cross section is also shown as the black dashed curve.
As seen in Fig. (\ref{fig:CrossSectionVsEnergyVsRatio}) the reactive
cross sections present a somewhat flat behavior whereas the elastic component is suppressed in the vicinity of 100 mK to 200 mK mostly due to the presence
of a resonant feature. However, it is worth noting that it may be premature
to consider this resonant structure as an actual observable feature
due in parts to the fact that our calculation only represents the
$J=0$ case. The incoherent summation of contributions associated to
higher $J$ values may (and are likely to) wash out these features observed
in Fig. (\ref{fig:CrossSectionVsEnergyVsRatio}). Thus, the question
whether the resonant structure predicted here shall survive the addition
of higher $J$ values will remain open for further theoretical explorations.
Likewise, the characterization of the resonance, in terms of angular
momentum partial waves, width and lifetime is outside the scope of
this work. However, as the entrance channel is associated to $j=0$
and $J=0$, and therefore only incoming $\ell=0$ partial wave contributes $\left(J=j+\ell\right)$, there is no centrifugal term associated
to the entrance channel potential curve, whose behavior is of an ordinary
attractive potential. As a consequence, it is possible that the resonance-like
feature shown in Fig. (\ref{fig:CrossSectionVsEnergyVsRatio}) is associated
to  a triatomic bound state belonging to another
channel, \textit{i.e.} a Feshbach resonance. This hypothesis is reinforced
by the somewhat high density of states that may exist in the vicinity of the entrance
channel -- see the red bold line in Fig. (\ref{fig:AphCoupledPotential}).

The hierarchy of the $j^{\prime}$-summed cross sections for a given
 $\upsilon^{\prime}$ level may be understood from the following considerations.
The $\upsilon^{\prime}=4$ manifold of LiF possess only 22 rotational
levels that are open with respect to the $\left(\upsilon=0,j=0\right)$ entrance
channel of CaF. This fact is illustrated, at 1 mK, in panel (a)
of Fig. (\ref{fig:OneMillikelvinAngularDistribution}). As a result,
the summation over $j^{\prime}$ yields the smallest cross sections
overall -- see the magnitude of the orange curve in Fig. (\ref{fig:CrossSectionVsEnergyVsRatio}). Similarly, the $\upsilon^{\prime}=3$ and 2 cases possess the second and third smallest amount of open rotational states (34 and 42), and thus, provides the second
and third smallest total cross section, \textit{i.e.} blue and green curves of Fig. (\ref{fig:CrossSectionVsEnergyVsRatio}). Overall the rotational levels of LiF belonging to the $\upsilon^{\prime}=4$-2 cases are predicted to be poorly populated by the collision in the energy range described here. In contrast, the production of LiF in the $\upsilon^{\prime}=0$ and 1 manifolds, the largest in terms of open rotational states,
are the chemical events with higher likelihood to occur, mostly populating $j^{\prime}=0$-20, with a smaller but substantial probability of populating also highly rotationally excited states.

Another noteworthy aspect, as the collision energy increases, is the somewhat strong suppression of the
elastic cross section that reaches a minimum value in the resonant region, at about 182 mK, as shown in the upper panel of Fig.
(\ref{fig:CrossSectionVsEnergyVsRatio}). As a consequence, in the range of energies
studied here, the Li + CaF collision may become predominantly reactive
at collision energies in the vicinity of 200 mK. This fact is illustrated by
the elastic-to-reactive ratio of the cross section presented in the
lower panel of Fig. (\ref{fig:CrossSectionVsEnergyVsRatio}), where
there are up to 1700 collisions for every chemical event at about
1 mK, remaining somewhat constant for about 100 mK, and quickly dropping to a minimum
of 1:1 (or smaller) in the vicinity of 182 mK. For
the sake of comparison, the elastic-to-inelastic ratio for a collision-induced
Zeeman relaxation of CaF by collisions with He atoms, at much higher
temperatures (2 K), has been measured by Maussang and co-workers to
be about 10$^{4}$ \cite{maussang2005}. Likewise, in the cases of
spin-polarized Li + CaH and Mg + CaH inelastic collisions,
investigated by Tscherbul \textit{et al.} \cite{tscherbul2011}, for
which  chemical reaction is energetically forbidden, the elastic-to-inelastic
ratio is predicted to be about 10$^{5}$ at 1 mK, \textit{i.e.} nearly 60 times
larger than the case considered here. In addition, CaH is known to be
more amenable for magnetic traps, mostly due to its higher rotational
constant compared to CaF \cite{lu2014}. This raises concerns
on the prospects of sympathetic cooling of CaF by means of cold
collisions with Li atoms above 100 mK due to potential
trap losses induced by the formation of LiF.

From an experimental point of view, our choice of using CaF in its lowest internal state as the entrance channel could be realized by producing the molecule with either a MOT or a Stark decelerator or a combination of these with a microwave trap. A modern MOT implementation is capable of producing molecules for collisions at energies as low as the Doppler limit whereas a Stark deceleration method is likely to produce molecules with a temperature of a few dozens of mK, being therefore more problematic for the case at hand. In either case, however, typical procedures, such as compressing the molecular cloud in order to improve its overlap with the buffer gas coolant may eventually raise the temperature by a few extra dozens of mK and, therefore, also trigger losses due to LiF formation. Those experimental implementation already reaching sub-Doppler temperatures should not be concerned by losses due to chemical reaction but sub-Doppler heating effects, as those demonstrated by Devlin and Tarbutt\cite{devlin2016}, may occur for certain kind of MOTs.

A detailed single-arrangement Lennard-Jones-based, and thus disregarding reactivity, simulation of the thermalization (no inelasticity either) of CaF in the presence of Li and Rb cold buffer gases has been carried out by Lim \textit{et al.} \cite{lim2015}. Their numerical experiment assumed a practical experimental scenario similar to that given above and found a somewhat strong dependency between the cooling rate and the $s$-wave scattering amplitude for those scenarios with ultracold Li atoms, when compared to Rb, mostly due to the relatively small reduced masses for the Li + CaF combination. In addition, they have also predicted a slowdown of the collision process, and thus the cooling rate, for the Li + CaF case, due to a minimum in the cross sections in the range of 1-10 mK, similar to that found in the present work at higher energies, about 100-200 mK. In contrast, a similar minimum was predicted by Lim \textit{et al.} within the $\mu$K range of collision energies when ultracold Rb atoms were used. As a consequence, the cooling rate when using Li was found to be an order of magnitude slower than that for the Rb case \cite{lim2015}.

Alternatively, as also pointed out by Lim \textit{et al.} \cite{lim2015}, the use of a light colliding partner such as Li for sympathetic cooling when CaF is produced in an excited rotational state may be favorable due to potentially higher centrifugal barriers that could, as a consequence, suppress losses due to collision-induced inelastic processes. This scenario is yet to be investigated but it is now possible using the PES we have presented here. Moreover, it may be worthwhile to explore collisions driven by the $^{3}A^{\prime}$ electronic state of the LiCaF complex and explore the possibility to control the reaction by means of external magnetic fields as in the case of the Li (Mg) + CaH systems \cite{tscherbul2011,singh2012,warehime2015}. Chemical reactions are likely to be suppressed in spin-polarized collisions of $^{2}S+{}^{2}\Sigma$ systems on the $^{3}A^{\prime}$ PES, due to the less attractive character of the $^{3}A^{\prime}$ electronic state and overall endothermicity. Moreover, there is evidence suggesting that the spin-orbit-induced triplet-to-singlet transition, that could trigger the formation of LiF in the $^{1}A^{\prime}$ PES, as shown here, may be either small or negligible \cite{warehime2015}. An overview of the $^{3}A^{\prime}$ electronic state of the LiCaF complex has been given by Frye and co-workers \cite{frye2016}.

Overall, our results show that cold collisions of Li and CaF favor elastic scattering in the 1-100 mK regime but a sharp decrease in the elastic cross section in the vicinity of 200 mK, possibly due to a Feshbach resonance, makes the elastic/reactive cross section ratio $<1$, limiting the efficacy of sympathetic cooling of CaF by collisions with cold Li atoms. Despite the high density of asymptotic diatomic states and bound triatomic states that are involved in the collisions -- see Fig. (\ref{fig:AphCoupledPotential}) --, our calculations predict a somewhat low density of resonances. This is probably due to the downhill nature of the reaction and presumably the short lifetimes of the LiCaF complexes formed. This aspect, the effect of rotational and vibrational excitation of the CaF molecule, and a proper characterization of the resonance will be addressed in future work. Indeed, a recent quantum close-coupling study of Ca + BaCl$^{+}$ system has shown strong vibrational quenching rates for BaCl$^{+}$ that exceeds rotational quenching rates for low-lying rotational levels~\cite{Stoecklin_2016}.

\section{Conclusions}
\label{sec:conclusions}

In this work we have applied state-of-the-art quantum chemistry and quantum reactive scattering to study both the interaction and dynamics of Li$\left(^{2}S\right)$ + CaF$\left(X^{2}\Sigma^{+}\right)$, in the context of cold collisions. To this end we have produced a global potential energy surface for the ground electronic state of the LiCaF system, $X^{1}$A$^{\prime}$, capable of describing both atom-diatom arrangements, Li + CaF and Ca + LiF. The electronic structure calculations were carried out using an internally contracted multi-reference configuration-interaction method with a state-averaged ($1^{1}$A$^{\prime}$, $1^{3}$A$^{\prime}$ and $1^{1}$A$^{\prime\prime}$) complete active space (10 active electrons in 9 active orbitals) self-consistent field electronic wavefunction. A total of about 11000 geometries were evaluated and used to produce the final potential energy surface fit with a many-body expansion method augmented with \textit{ab initio} parameterized long-range potentials.

Scattering calculations for the Li + CaF$(\upsilon=0,j=0)$ entrance channel were performed between 1 and 200 mK of collision energy. At 1 mK the collision-induced formation of rovibrationally excited LiF$(\upsilon^{\prime}=0$-1, $j^{\prime}\approx0$-20$)$ molecules is predicted to be the most likely collisional outcome, with a total energy release that could reach up to 3880 K. In the vicinity of 100-200 mK a quantum resonance, likely to be a Feshbach resonance, appears to strongly suppress the elastic component. The reactive cross sections, however, remain largely unaffected in this regime, presumably due to its small magnitude compared to its elastic counterpart. The overall effect is that the elastic-to-reactive cross section ratio falls well below the lower limit of one hundred for collision energies above 100 mK suggesting a somewhat poor cooling rate for sympathetic cooling of CaF by Li and a strong trap loss due to the formation of LiF. At the resonance energy of 182 mK nearly every collision is predicted to be reactive (1:1 ratio or smaller).

It is worthwhile to emphasize, however, that the calculations presented here are not yet accurate for direct comparisons with future experimental observations as most likely a single PES, and the single partial wave $(J=0)$ used in the scattering calculations are insufficient. However, we believe it will serve as a benchmark for further theoretical works as we provided a detailed description of the potential energy surface and of those numerical aspects required to obtain reasonably well converged scattering characteristics, a substantial improvement upon previous studies that were limited to model potentials and elastic/inelastic collisions, oftentimes considering only the equilibrium geometry of the triatomic complex, and equally limited dynamical models.

\section*{Author Contributions}
Electronic structure calculations were primarily carried out by Q.Y. and H.G. Scattering calculations were carried out by H.S. with assistance from M.M., N.B. and B.K.K. All authors contributed to manuscript preparation and editing.

\section*{Conflicts of interest}
There are no conflicts to declare.

\section*{Acknowledgements}
This work is supported in part by NSF grant No. PHY-2110227 (N.B.) and by a MURI grant from Army Office of Research (Grant No. W911NF-19-1-0283 to H.G. and N.B.). The computation was performed in part at the Center for Advanced Research Computing (CARC) at UNM, and used the Extreme Science and Engineering Discovery Environment (XSEDE), which is supported by the National Science Foundation (Grant No. ACI-1548562). Specifically, it used the Bridges-2 system, which is supported by the NSF (Award No. PHY-200034) (N.B.) at the Pittsburgh Supercomputing Center (PSC). B.K.K. acknowledges that part of this work was performed under the auspices of the US Department of Energy under Project No. 20170221 ER of the Laboratory Directed Research and Development Program at Los Alamos National Laboratory. This work used resources provided by the Los Alamos National Laboratory Institutional Computing Program. Los Alamos National Laboratory is operated by Triad National Security, LLC, for the National Nuclear Security Administration of the U.S. Department of Energy (Contract No. 89233218CNA000001).



\balance


\bibliography{rsc} 
\bibliographystyle{rsc} 

\end{document}